\documentclass[final,5p,times,twocolumn,lefttitle]{elsarticle} 
\usepackage{graphicx}%
\usepackage{multirow}%
\usepackage{amsmath,amssymb,amsfonts}%
\usepackage{amsthm}%
\usepackage{mathrsfs}%
\usepackage[title]{appendix}%
\usepackage{xcolor}%
\usepackage{textcomp}%
\usepackage{manyfoot}%
\usepackage{booktabs}%
\usepackage{algorithm}%
\usepackage{algorithmicx}%
\usepackage{algpseudocode}%
\usepackage{listings}%

\usepackage{orcidlink}
\usepackage{tabularx}
\usepackage{siunitx}
\usepackage{newtxtext,newtxmath}
\usepackage{microtype}
\hypersetup{colorlinks=true,allcolors=blue}
\newcommand{\loss}{{G^{\prime\prime}}}
\newcommand{\elastic}{{G^{\prime}}}
\newcommand{\sfc}{\sigma_y^{\rm FC}}
\newcommand{\sosc}{\sigma_y^{\rm osc}}

\newcommand{\sStep}{\sigma_y^{\rm step}}
\newcommand{\figref}[1]{Fig.\,\ref{#1}}
\newcommand{\partfigref}[2]{Fig.\,\hyperref[#1]{\ref*{#1}(#2)}}
\newcommand{\secref}[1]{Sec.\,~\ref{#1}}
\newcommand{\refcite}[1]{Ref.\,\cite{#1}}
\newcommand{\phip}{\phi_{\rm alp}}
\newcommand{\phim}{\phi_{\mu}}
\newcommand{\phicp}{\phi_{\rm rcp}}
\newcommand{\ie}{\textit{i.e.}}
\newcommand{\cf}{\textit{c.f.}}
\newcommand{\eg}{\textit{e.g.}}

\journal{arXiv}
\begin{document}

\renewcommand{\floatpagefraction}{.8}%

\begin{frontmatter}
\title{Fresh Cement as a Frictional Non-Brownian Suspension}

\author{{James A. Richards} \orcidlink{0000-0002-2775-6807}\corref{cor}}\ead{james.a.richards@ed.ac.uk}
\author{Hao Li\orcidlink{0009-0001-8922-4830}}
\author{{Rory E. O'Neill} \orcidlink{0000-0002-4901-7087}}\ead{rory.o'neill@ed.ac.uk}
\author{{Fraser H.\ J. Laidlaw}\orcidlink{0000-0002-5907-0447}}
\author{{John R. Royer} \orcidlink{0000-0002-8368-7252}}\ead{john.royer@ed.ac.uk}
\address{Edinburgh Complex Fluids Partnership and School of Physics and Astronomy, The University of Edinburgh, James Clerk Maxwell Building, King's~Buildings, Edinburgh EH9 3FD, United Kingdom}

\cortext[cor]{Corresponding author}

\begin{abstract}
Cement is an essential construction material due to its ability to flow before later setting, however the rheological properties must be tightly controlled. Despite this, much understanding remains empirical. Using a combination of continuous and oscillatory shear flow, we compare fresh Portland cement suspensions to previous measurements on model non-Brownian suspensions to gain a micro-physical understanding. Comparing steady and small-amplitude oscillatory shear, we reveal two distinct jamming concentrations, $\phi_{\mu}$ and  $\phi_{\rm rcp}$, where the respective yield stresses diverge. As in model suspensions, the steady-shear jamming point is notably below the oscillatory jamming point, $\phi_{\mu} < \phi_{\rm rcp}$, suggesting that it is tied to frictional particle contacts. These results indicate that recently established models for the rheology of frictional, adhesive non-Brownian suspensions can be applied to fresh cement pastes, offering a new framework to understand the role of additives and fillers. Such micro-physical understanding can guide formulation changes to improve performance and reduce environmental impact.
\end{abstract}


\end{frontmatter}

\section{Introduction}\label{sec1}

Concrete is the most used construction material, with global annual production over \SI{30}{\giga\tonne}~\cite{miller2020climate}, or over three tonnes per capita~\cite{gagg2014cement}. The multitude of applications is driven by the set mechanical properties, \eg, high compressive strength~\cite{li2020durability}, combined with the ability to transport, process and form in a ``fresh'' fluid state. Basic fresh concrete consists of a mix of aggregates (sands and gravel $\gtrsim \SI{1}{\milli\metre}$) in a background slurry of Portland cement with water. This slurry is a suspension of finer particles from \SI{100}{\micro\metre} down to $\lesssim \SI{1}{\micro\metre}$~\cite{bentz1999effects}, which partially dissolves before reacting with water to set and bind the aggregates~\cite{ioannidou2016crucial}.

The flow of fresh concrete before the initial setting time of \SI{45}{\minute} to \SI{6.5}{\hour}~\cite{zhang2011tests} is dominated by the properties of this cement background, except at the highest aggregate proportions~\cite{yammine2008ordinary}. Even at very high aggregate loadings the cement slurry rheology can play an outsized role under certain flow conditions, for example when pumped through pipes the resistance is determined by the rheology of an aggregate-free layer near the wall~\cite{choi2013lubrication}. Thus, understanding and controlling the cement slurry rheology is key for optimising the flow of the overall concrete mix.

Specifically, the yield stress of this cement background controls number of key aspects during the handling and forming of fresh concrete and is subject to duelling requirements. When poured, the yield stress must be low enough to enable steady flow and compaction. In contrast, following pouring the yield stress must be large enough ensure aggregates to remain mixed~\cite{saak1999characterization}. In this stage a larger yield stress is also desired to reduce pressure on the formwork holding the concrete in place~\cite{ovarlez2006physical}. With the growing usage of concrete-based 3D printing these competing constraints are magnified, with concrete needing to flow through a nozzle before retaining its extruded shape under its own weight and then of layers above~\cite{roussel2018rheological}. Therefore, understanding the flow and yield stress of the cement background is vital for performance.

Portland cement is the primary contributor to concrete's environmental impact, accounting for 8\% of global carbon dioxide emissions~\cite{miller2020climate}, principally from calcining and fusing limestone and clay into clinker that is then ground~\cite{worrell2001carbon}. Reducing Portland cement use via replacement binders requires retaining both set and flow properties~\cite{panesar2020performance}. The rheology of fresh cement with alternative materials has, therefore, been extensively tested~\cite{ferraris2001influence,jiang2020utilization,wu2019changes,ting2019effects}. Prediction of how these replacements change the rheology is still developing. The challenge arises from cement suspensions lying between two distinct regimes. For larger granular particles, such as the aggregates, interactions have well-defined contact~\cite{boyer2011unifying} and solid friction~\cite{guazzelli2018rheology}. In contrast, sub-micrometre colloidal particle interactions are controlled by surface-chemistry--derived potentials~\cite{royall2013search} and random close packing, $\phicp$, the jamming point for frictionless particles~\cite{silbert2010jamming}.

The intermediate ``non-Brownian'' regime has recently been shown to be controlled by a delicate interplay of friction and inter-particle potentials~\cite{guy2015towards,clavaud2017revealing}. At high solid volume fraction, this leads to non-Newtonian behaviour, such as shear thickening from the applied stress overcoming repulsive inter-particle forces leading to contact~\cite{lin2015hydrodynamic,comtet2017pairwise}. Shear thickening is then controlled by a frictional jamming point~\cite{wyart2014discontinuous} lower than $\phicp$~\cite{silbert2010jamming}. This paradigm has been extended to different interaction potentials~\cite{guy2018constraint,richards2020role} and transient rheology~\cite{richards2019competing,han2019stress}.

Here we show from comparing multiple rheological tests that the flow of a fresh Portland cement suspension is controlled by a jamming volume fraction below random close packing, which we associate with friction. The flow can then be compared to non-setting model suspensions. We first present flow properties over a wide range of volume fractions under different shear protocols. From this we identify two distinct yield stresses: firstly, from shear in a single continuous direction, \ie\ anisotropic; secondly, after preparation via a decreasing strain amplitude sweep with no preferred direction, or ``isotropic''. Comparing these yield points, and how they are connected by strain, enables the role of friction to be disentangled. Identifying the import of friction has several ramifications, for example, in considering the setting reaction under flow, \ie\ thixotropy, and how the background rheology interacts with the aggregate. It also influences how the maximum packing fraction relates to particle morphology, relevant to formulation and optimisation of cement alternatives and admixtures.

\section{Materials and Methods\label{sec:mater}}

\subsection{Preparation and characterisation}

\begin{figure}
    \centering
    \includegraphics[width=\columnwidth]{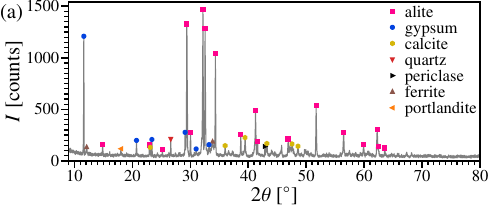}
    \includegraphics[width=\columnwidth]{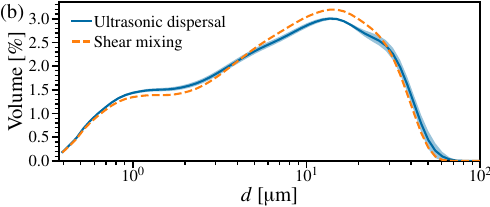}
    \caption{Cement particle characterisation. (a)~X-ray powder diffraction, intensity count (red) as a function of diffraction angle, $2\theta$. Matched peaks, see inset legend for colours. Primary component, alite (calcium trisilicate); secondary components, limestone (calcium carbonate, note that the 100\% peak overlaps a peak of alite, and it is instead identified by secondary peaks) and gypsum (calcium sulphate dihydrate); minor components, periclase (magnesium oxide), quartz (silicon dioxide), portlandite (calcium hydroxide) and ferrite (calcium aluminoferrite). (b)~ Particle size distribution (PSD) for cement powder. Volume weighted distribution of diameters, $d$, from laser diffraction at 25 bins/decade. Lines: solid, PSD for ultrasonic dispersal with shading standard deviation over three samples ($d_{10} = \SI{1.0}{\micro\metre}$, $d_{50} = \SI{7.2}{\micro\metre}$, $d_{90} = \SI{27.3}{\micro\metre}$); dashed, PSD for cement dispersed by mixing at high solids concentration ($d_{10} = \SI{1.0}{\micro\metre}$, $d_{50} = \SI{7.5}{\micro\metre}$, $d_{90} = \SI{26.0}{\micro\metre}$).}
    \label{fig:psd}
\end{figure}

The cement [CEM-II/A-LL 32,5R Portland-composite cement (Blue Circle, Tarmac CRH), BS EN 197-1] is prepared with varying water-to-cement ratios ($w/c$) from 0.23 to 0.50. Alongside clinker as the majority component, the presence of minority non-setting, components of limestone (CEM-II specifying a 6--20\% content), and gypsum (to delay initial setting) was confirmed using X-ray powder diffraction [Rigaku SmartLab, Bragg Brentano geometry, Cu K$\alpha_1$ radiation (wavelength \SI{0.15401}{\nano\metre}) from sealed tube (\SI{40}{\kilo\volt}, \SI{50}{\milli\ampere}) with Johannsson monochromator, HyPix3000 detector (1D scan mode) and incident/receiving soller slits \SI{2.5}{\degree}. Scan parameters: scattering angle, $2\theta = \SIrange{9}{80}{\degree}$, step size \SI{0.01}{\degree} at \SI{0.8 }{\degree\per\minute}, length limiting slit of \SI{10}{\milli\metre}, and incident slit of \SI[parse-numbers = false]{1/3}{\degree}.], \partfigref{fig:psd}{a}~\cite{stutzman2016phase}. 

The dispersed particle size distribution (PSD) was measured using laser diffraction (Beckman Coulter LS 13 320). A 5\,wt\% stock suspension in isopropyl alcohol was prepared via equilibration on a roller bank for \SI{30}{\minute} and ultra-sonic probe dispersal (\SI{200}{\watt}, \SI{5}{\second} pulses for \SI{9}{\minute} total), based on \refcite{hackley2004particle}. Three samples were measured from the stock suspension via dilution to 1\,wt\% and mixing to 10\% attenuation, with three PSD measurements of \SI{30}{\second} to assess both measurement and sample variability. The cement has a broad bimodal PSD of diameter $d = \SIrange{0.375}{70}{\micro\metre}$ with fines from \SIrange{1}{2}{\micro\metre} and a primary population around \SI{14}{\micro\metre}, \partfigref{fig:psd}{b}~[solid (blue) line].

\begin{figure*}
    \centering
    \includegraphics[width=\textwidth]{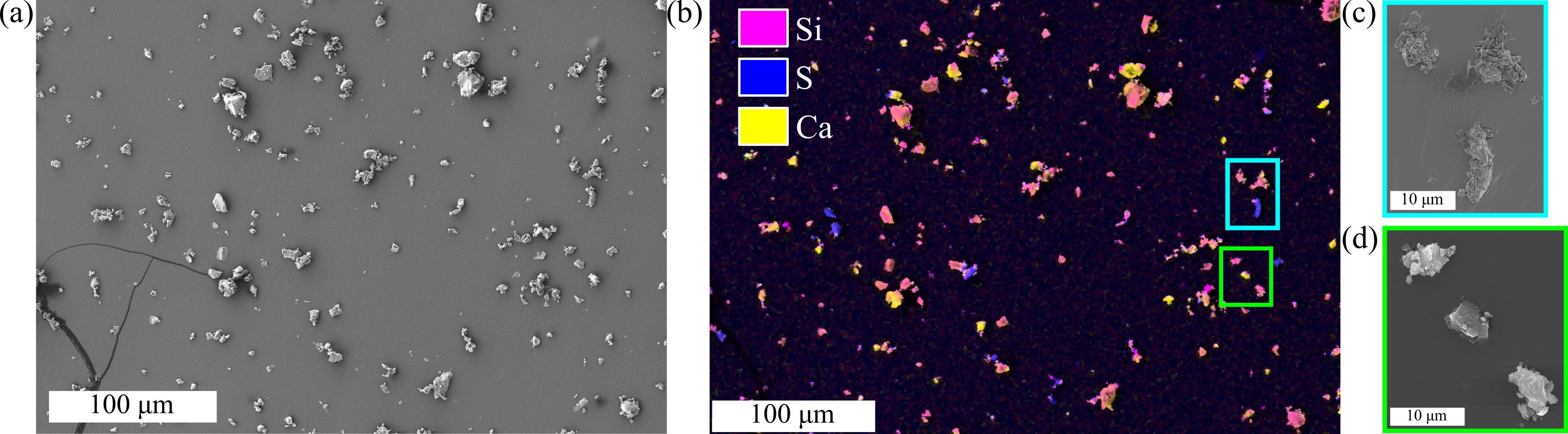}
    \caption{Energy dispersive X-ray mapping of a dry cement powder. (a)~Secondary electron image of spray dispersed dry powder, \SI{100}{\micro\metre} scale bar inset. (b)~Elemental mapping of corresponding area, silicon (Si, pink); sulphur (S, blue); and calcium (Ca, yellow). (c)~Magnified secondary electron image of upper (blue) highlighted region comparing cement and gypsum, \SI{10}{\micro\metre} scale bar inset. (d)~Equivalent comparison of cement and calcite in lower (green) highlighted region.}
    \label{fig:edx}
\end{figure*}

The composition was verified using scanning electron microscopy (SEM) of a dispersed dry powder (Zeiss Crossbeam 550, powder dispersed via compressed air onto carbon tape and platinum coated, \SI{5}{\kilo\eV} acceleration voltage), \partfigref{fig:edx}{a}. Energy dispersive X-ray imaging (Oxford Instruments X-Max$^{\rm n}$ 150, \SI{20}{\kilo\eV} with \SI{500}{\pico\ampere} current) maps varying concentrations of elements across the sample, \partfigref{fig:edx}{b}; we overlay silicon (pink), sulphur (blue) and calcium (yellow). Pink then highlights the majority clinker component (primarily calcium trisilicate), blue the gypsum, and yellow (\ie\ the absence of Si and S) calcite. Imaging of selected regions comparing cement particles with gypsum, \partfigref{fig:edx}{c}, and calcite, \partfigref{fig:edx}{d}, illustrates that the various components have similar particle sizes of $\mathcal{O}(\SI{10}{\micro\metre})$ and a faceted morphology. This justifies our rheological treatment of them as a homogeneous blend of non-Brownian particles. Traces of iron, aluminium, sodium and potassium were dispersed throughout the sample, with magnesium localised to small regions~\cite{song2021occurrence}.

Rheological samples were prepared through adding distilled water to \SIrange{5}{10}{\gram} of cement powder by spatula mixing. To achieve a homogeneous paste, free of visible powder agglomerates, the suspension was vortex mixed for \SI{1}{\minute}; preparation and loading was $\leq \SI{10}{\minute}$. To verify agglomerate break down, a sample was prepared using isopropyl alcohol equivalent to a $w/c = 0.38$ followed by dilution to 2\,wt\%. A comparable PSD was measured with equivalent method, cf.~dashed and solid lines, \partfigref{fig:psd}{b}, with $d_{90} = \SI{26.0}{\micro\metre} \approx \SI{27.3}{\micro\metre}$.

\begin{figure}[t]
    \centering
    \includegraphics[width=\columnwidth]{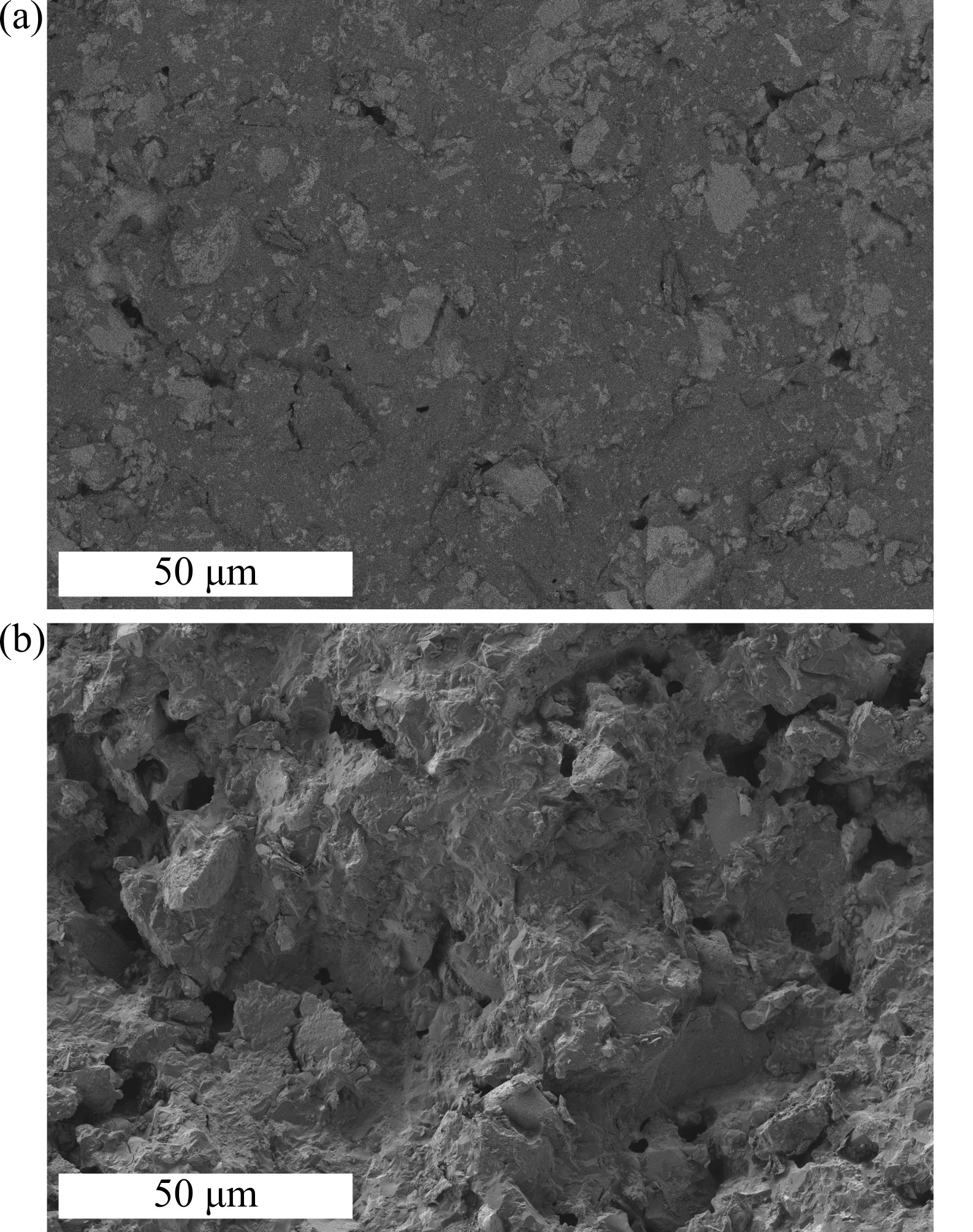}
    \caption{Cryogenic scanning electron microscopy of freeze-fractured cement suspension \SI{10}{\minute} after mixing. (a)~energy selective backscattering, particles (light) and water background (dark), scale bar \SI{50}{\micro\metre}. (b)~Corresponding secondary electron topographic image.}
    \label{fig:sem}
\end{figure}

To confirm the composition and structure when mixed, cryogenic-SEM of a $w/c = 0.45$ sample after \SI{10}{\minute} was performed (Quorum Technologies Ltd PP3010T cryogenic attachment; samples freeze-fractured using liquid-nitrogen slush at \SI{-140}{\celsius} with ice sublimed at \SI{-90}{\celsius} for \SI{5}{\minute}). Using uncoated samples with a \SI{1.2}{\kilo\eV} acceleration voltage, energy selective backscattering (\SI{700}{\volt} filter) imaging, \partfigref{fig:sem}{a}, highlights the dispersed particles (light) in the water background (dark). The corresponding topographic image is given by a secondary electrons secondary ions detector, \partfigref{fig:sem}{b}. This confirms the presence of primary $\mathcal{O}(\SI{10}{\micro\metre})$ particles with a significant fraction of micrometre-sized fines in a dispersed state. 

We report results in terms of an apparent solid volume fraction, $\phi = 1/(\rho_c w/c +1)$, where $\rho_c = 3.15$, a literature specific gravity of a Portland cement particles~\cite{flatt2004dispersion}. The maximum dry packing fraction can then be measured as 0.68(1) from compacted powder density using a Rigden apparatus~\cite{rigden1947use}. 

\subsection{Loading and pre-shear}

As we focus on the cement suspension alone, conventional geometries with $\mathcal{O}(\SI{1}{\milli\metre})$ gap sizes can be used without the confinement effects which become problematic  with suspensions containing larger aggregates. A cross-hatched parallel-plate geometry (radius $R = \SI{20}{\milli\metre}$, TA Instruments DHR-2 with Peltier plate at \SI{20}{\celsius}) was used to reduce wall slip, a known issue in cement rheology~\cite{vance2015rheology}. Samples were loaded by lowering the upper plate in steps until the sample filled the gap, any small excess was trimmed. Though this lead to a varying gap, typically $h = \SIrange{1.2}{1.8}{\milli\metre}$, this was necessary to avoid high-pressure squeeze-flow, and the associated liquid migration~\cite{delhaye2000squeeze} that can change the composition in cement-based materials~\cite{grandes2021rheological}. Tests were conducted under imposed strain, utilising the feedback loop of the stress-controlled rheometer, with rim strain, $\gamma = \Theta R/h$, from angular displacement, and apparent stress, $\sigma = 2\mathcal{T}/\pi R^3$, from torque.

Before each test the sample was prepared with a decreasing oscillatory strain amplitude sweep from $\gamma_0 = 0.3$ to $10^{-4}$ at angular frequency $\omega = \SI{5}{\radian \per \second}$ with 5 pts/decade (6 for $w/c < 0.30$). Each point equilibrated for two periods, before measurement from correlation of four cycles (point time $\approx \SI{8}{\second}$). Pre-shear reduces the varying influence of sample history, with alternate paths through mixing and loading. Typically, samples are pre-sheared at high rate~\cite{raghavan1995shear}. While this effectively equalises the shear history, in a concentrated suspension this may drive migration and inhomogeneity~\cite{lecampion2014confined}. This is seen in Couette cylinders, with migration away from the inner cylinder~\cite{fall2010shear}. Similarly, for a parallel-plate an equivalent region of lower shear rate exists towards the centre. Additionally, with cross-hatching the sample within the serrations is subject to a lower shear rate~\cite{carotenuto2015predicting}.

\subsection{Rheometric protocols}

Each sample was measured with an oscillatory test, followed by a continuous (single direction) shear test. The oscillatory test was an increasing amplitude sweep from $\gamma_0 = 10^{-4}$ to 1, with the same settings as the preparation step. We report the elastic, $\elastic$, and loss, $\loss$, moduli from the first Fourier components. The continuous shear tests comprised stress growth at a fixed shear rate $\dot\gamma = \SI{0.05}{\per\second}$ for \SI{120}{\second}, followed by an increasing flow sweep from \SIrange{0.05}{50}{\per\second} at 5 pts/decade (6 for $w/c < 0.30$) with \SI{5}{\second} equilibration and \SI{5}{\second} measurement per point. Tests were performed in this order as the flow sweep may lead to sample fracture. Multiple repeat samples were prepared and measured following this same protocol, with a between 3 to 5 repeats for $w/c \geq 0.27$ ($\phi\leq 0.54$), see Table~\ref{tab:repeats}. Due to challenges in loading at low $w/c$, only two samples were measured for 0.25 and one at 0.23. Although a paste could be prepared at $w/c = 0.21$, it was not possible to uniformly load.

These protocols were based on previous measurements of non-Brownian yield-stress fluids~\cite{richards2020role,richards2021turning}, with the density of measurement points and the time per point reduced. The reduced overall test time ($\approx\SI{20}{\minute}$ from mixing to completion) allowed us to avoid changes in the sample composition due to solvent evaporation and sedimentation, which were more acute using water as a suspending medium compared to lower volatility and higher viscosity oils or glycerol-water mixtures used in other model systems. This also ensured measurements were completed before effects of the initial setting reaction become appreciable (over a timescale of $\approx$ hours~\cite{zhang2011tests}). Despite these protocol adjustments, sedimentation prevented us from working with lower volume fraction samples $\phi<0.38$, as the more rapid sedimentation resulted in visible separation during pre-shear. 

\section{Results}\label{sec:res}

We first outline the rheology with changing $\phi$ in the two main tests, continuous flow curve and oscillatory strain amplitude sweep, followed by step shear. We focus on identifying yielding events. In \secref{sec:dis}, we then compare and link these to probe micro-physics.

\subsection{Flow curve measurements\label{sec:res:fc}}

\begin{figure}
    \centering
    \includegraphics[width=\columnwidth]{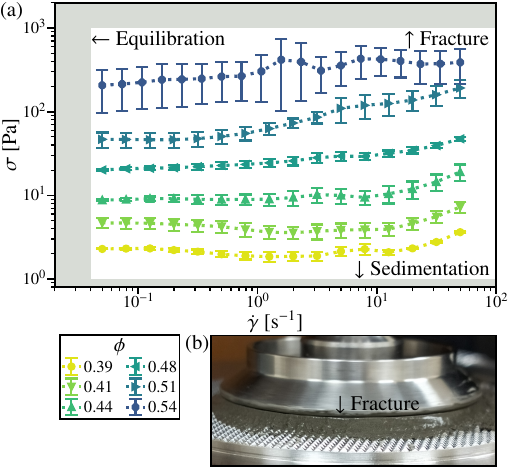}
    \caption{Steady-shear rheology of cement suspensions with varying solid volume fraction. Points, stress as a function of applied shear rate, $\sigma(\dot\gamma)$, at given solid volume fraction, $\phi$ (see inset legend). Error bars, standard deviation in repeat measurements. Shading, inaccessible measurement regions with labelled limitations: maximum $\sigma$, sample fracture; minimum $\sigma$, sedimentation stress scale; and minimum $\dot\gamma$ limited by equilibration (accumulation of $\gamma \approx 6$ without sedimentation or drying). Inset: image of fracture for sample at $\phi>0.54$.}
    \label{fig:fc}
\end{figure}

Measurement of a continuous flow curve of stress with shear rate, $\sigma(\dot\gamma)$, is a common characterisation for concrete. This potentially stems from its similarity to applications~\cite{ferraris2001influence}, for example, pumping through a pipe. For a cement suspension, with increasing $\phi$ the measured stress trends upwards, \partfigref{fig:fc}{a} light (yellow) to dark (purple) symbols. For a given $\phi$, the stress changes weakly with shear rate. For all $\phi$ measured (corresponding to $\phi \leq 0.54$), at the lowest shear rates the stress becomes independent of applied rate, indicative of a yield stress.

We extract a flow curve yield stress, $\sfc$, taking the average stress over the rage $\dot\gamma = \SIrange{0.05}{0.2}{\per\second}$ where $\sigma\approx {\rm \ constant}$. This yield stress increases steadily with $\phi$ from $\SI{2}{\pascal}$ at $\phi = 0.39$ to $\SI{50}{\pascal}$ at $0.51$. Above this, at $\phi = 0.54$, \partfigref{fig:fc}{a}~(dark purple), $\sfc$ increases sharply to $\sim \SI{200}{\pascal}$ with increased sample variability.  In contrast to the  measurements at lower $\phi$, we find a weak but notable gap-dependence, with higher measured $\sfc$ at lower $h$. This suggests that the measured $\sfc$ at this point is not a well-defined material parameter, likely an underestimate of the true steady-shear yield stress due to inhomogeneous flow. At higher $\phi$ still, although loadable, the sample fractures instead of flowing, with a clear fracture plane near the top plate throughout the sample, \partfigref{fig:fc}{b}.

At higher shear rates, up to $\dot \gamma \approx \SI{50}{\per\second}$, we observe a weak increase in the shear stress $\sigma(\dot\gamma)$, though for some of our lowest volume fractions $\sigma(\dot\gamma)$ first decreases before increasing. These samples do not strictly behave as ``simple'' yield stress fluids~\cite{ovarlez2013existence}, implying an element of thixotropy (decrease in stress with time), oft associated with shear banding. In contrast, at our highest concentrations $\sigma(\dot\gamma)$ initially increases monotonically but then plateaus at our highest shear rates. While such behaviour can be associated with wall-slip~\cite{meeker2004slip_a,meeker2004slip_b}, we regard this as unlikely in our cross hatched plates and instead suspect either inhomogenous flow or edge fracture. Indeed, beyond $\dot \gamma \approx \SI{50}{\per\second}$ sample fracture or other edge instabilities, such as ejection of material from the gap, become visibly evident. This suggests that non-rheometric effects become significant even before this point, and thus do not further probe this high shear-rate regime, \eg, fitting a Bingham model to extract a viscosity.

To access a robust high shear viscosity an alternate geometry would be required, \eg, a Couette cylinder. However, loading high $\phi$ samples in such geometries is exceptionally challenging and would preclude our main focus (though fractal vane geometries may provide a promising future route~\cite{owens2020improved,chaparian2022computational}). 

\subsection{Oscillatory shear response\label{sec:res:osc}}

\begin{figure}
    \centering
    \includegraphics[width=\columnwidth]{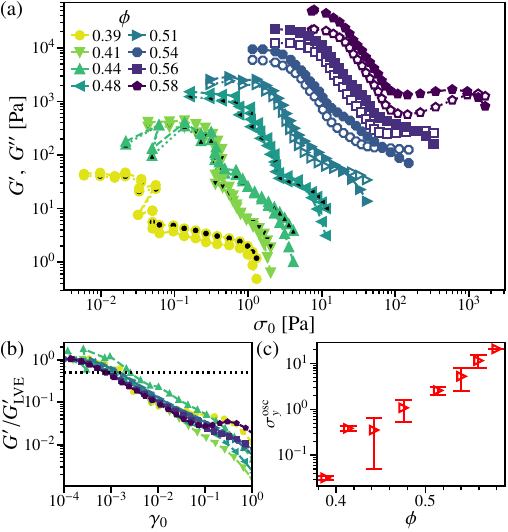}
    \caption{Oscillatory rheology. (a)~Elastic modulus ($G^{\prime}$, filled symbols) and loss modulus ($G^{\prime\prime}$, open) with oscillatory stress, $\sigma_0 = \gamma_0(\elastic^2+\loss^2)^{1/2}$, from increasing strain amplitude sweep, $\gamma_0$, at increasing volume fraction, $\phi$, light (yellow) to dark (blue), see legend for values. 
    (b)~Normalised elastic modulus, $\elastic/\elastic_{\rm LVE}$, vs strain $\gamma_0$. We compute $\elastic_{\rm LVE} $ averaging the first three points, and define yielding where $\elastic/\elastic_{\rm LVE}=0.5$ (dashed line). (c)~Corresponding oscillatory yield stress, $\sosc$, with $\phi$.}
    \label{fig:osc}
\end{figure}

While most work on cement rheology focuses on steady flow, as this is most relevant for transport and processing, oscillatory tests can be used to probe the attractive particle interactions and build up of a gel-like network. Prior work using small amplitude shear examined how the linear viscoelastic (LVE) response  changed with varying composition~\cite{liberto2022small} or during setting~\cite{bellotto2013cement,nachbaur2001dynamic}. Here we probe with changing oscillation strain amplitude to identify an oscillatory yield stress, $\sosc$, across a broad $\phi$ range. 

The elastic ($\elastic$, filled symbols) and loss ($\loss$, open) moduli for samples at varying $\phi$ are plotted as a function of oscillation stress, $\sigma_0 = \gamma_0 \sqrt{\elastic ^2 + \loss ^2}$, \partfigref{fig:osc}{a}. These tests could be performed without sample fracture up to $\phi =0.58$, in contrast to the flow curves where fracture occurred for $\phi \gtrsim 0.54$. At a given strain the stress response increases with volume fraction, as with $\sigma(\dot\gamma)$, so that both moduli shift to the upper right when plotted against $\sigma_0$. 

Qualitatively, samples at all $\phi$ show the same behaviour, with a linear viscoelastic response at small $\gamma_0$, then yielding as the strain amplitude increases. In the LVE region ($\gamma_0 \lesssim 3 \times 10^{-4}$) the moduli are strain independent and $\elastic > \loss$, indicating the cement pastes are solid-like over the timescale $2\pi/\omega \sim \SI{1}{\second}$. 

With increasing $\gamma_0\gtrsim 10^{-3}$ the moduli drop, indicating yielding~\cite{dinkgreve2016different} and the onset of plastic rearrangements~\cite{richards2021characterising}. While the drop in moduli appears gradual against $\gamma_0$, \partfigref{fig:osc}{b}, it is rather abrupt in terms of  $\sigma_0$.  This is particularly evident at our lower concentrations, $\phi \lesssim 0.44$, where the moduli, though noisy, drop by an order of magnitude or more at a critical stress. With increasing $\phi$ this yielding becomes less abrupt, although the moduli still drop up to a hundred-fold over a narrow stress window. We define the oscillatory yield stress, $\sosc$, by a 50\% reduction in $\elastic$ relative to the LVE plateau, though our conclusions are robust to the precise threshold, \eg\ a five-fold reduction alters the precise values but not the overall behaviour of $\sosc(\phi)$.

The extracted $\sosc$ increases with $\phi$, \partfigref{fig:osc}{c}, from $\sosc\approx\SI{0.1}{\pascal}$ at our lowest $\phi$ up to $\sosc\approx\SI{20}{\pascal}$ at $\phi=0.58$. Note that this oscillatory yield stress is finite at about \SI{5}{\pascal} for $\phi = 0.54$, the highest concentration where we achieve steady flow without fracture, with $\sosc(\phi)$ increasing more rapidly beyond this point. 

Moving through the yielding transition, the loss modulus $\loss$ drops along with $\elastic$ as $\gamma_0$ increases. This is distinct from yielding in typical soft solids, \eg, a jammed emulsion, where $\loss$ either first rises or decreases more slowly (see the Type III and Type I behaviours, respectively, classified in~\cite{hyun2011review}). 
At higher $\gamma_0$ the moduli develop a shoulder, with the drop slowing (or even reversing at high concentrations), before decreasing again. Only here, in this highly non-linear regime, do the moduli cross, rendering the typical operative $\elastic = \loss$ yielding definition questionable.

The oscillatory yielding point appears more sharply defined in a {\it decreasing} oscillatory strain amplitude sweep, with $\elastic(\sigma_0)$ displaying a re-entrant `nose', see Appendix \partfigref{fig:oscDown}{a--b}. Nearly identical behaviour was previously observed in oscillatory down sweeps with a model adhesive non-Brownian suspension (cornstarch in oil)~\cite{richards2020role}.
While the precise form of $\elastic(\sigma_0)$ differs between the up- and down-sweeps, the oscillatory yield stresses obtained from two sets of curves closely agree at all but the highest $\phi \geq 0.56$, \partfigref{fig:oscDown}{c}, where the down-sweep yield stress plateaus around \SI{5}{\pascal} while the up-sweep yield stress continues to rise. At these highest concentrations the down-sweep rheology is extremely sensitive to the maximum strain amplitude, and such a plateau likely reflects an unnoticed sample fracture rather than reflecting the true bulk rheology. We thus focus on the $\sosc$ measured from the up-sweeps in our later discussion and comparison with $\sfc$ measured from the flow curves, though note that our overall conclusions are insensitive to this choice.

\subsection{Start-up shear\label{sec:res:step}}

\begin{figure}
    \centering
    \includegraphics[width=\columnwidth]{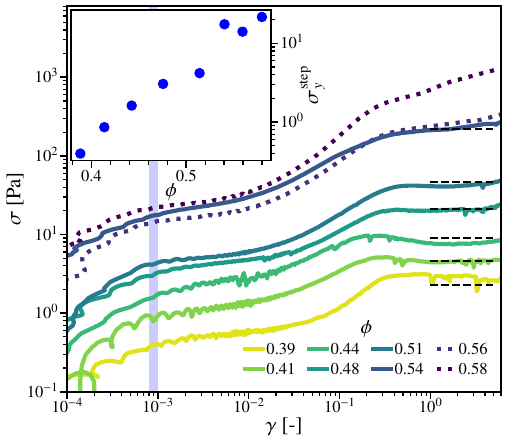}
    \caption{Step shear response after oscillatory preparation. Stress vs strain, $\sigma(\gamma)$, at fixed shear rate $\dot\gamma = \SI{0.05}{\per\second}$ with increasing volume fraction $\phi$, light (yellow) to dark (blue), see legend. Lines: solid, samples without visible fracture at end of step; dashed, corresponding $\sfc$; and, dotted, samples with visible fracture. Shading, $0.0008 \leq \gamma \leq 0.001$ defining step yield stress, $\sStep$, $\phi$ dependence inset.}
    \label{fig:step}
\end{figure}

In between the oscillatory testing and  flow curve measurement (\secref{sec:res:fc}), we apply a fixed-rate step shear up to $\gamma=6$ at our lowest probed steady shear rate $\dot\gamma = \SI{0.05}{\per\second}$. This step ensures we reach a steady state prior to our flow curve acquisition, as inadequate equilibration at low rates can result in a spurious low-shear viscosity~\cite{dinkgreve2017everything} or underestimation of the yield stress. The strain dependent dynamics during this step shear also explicitly links the small-strain state (\ie\ oscillatory test) and the large-strain flow curve. 

At low strains, $\gamma < 5 \times 10^{-4}$, the stress rises nearly linearly, \figref{fig:step}, corresponding to a solid-like elastic response for a solid, though detailed characterisation of this small-strain regime is limited by the response time of our stress-controlled rheometer. With increasing $\gamma$ from $10^{-3}$ to $10^{-2}$, the stress turns over and plateaus, a standard indication of yielding in a shear start-up test~\cite{dinkgreve2016different}. We can thus define a step shear yield stress, $\sStep$, from the average stress in this cross-over region (indicated by blue shading). As $\phi$ increases over our measured range, $\sStep$ increases from \SIrange{0.5}{20}{\pascal} (inset). 

In conventional soft solids, this stress-plateau would continue for larger strains~\cite{dinkgreve2016different}. However, for our fresh cement suspensions we instead find pronounced strain hardening from $\gamma \geq 0.01$ to 0.3--0.5. For $\phi \leq 0.54$, $\sigma(\gamma)$ again plateaus at large $\gamma \geq 1$ indicating a second yielding event; this corresponds to the flow curve yield stress, \textit{cf.}\ solid and dashed lines, \figref{fig:step}~[yellow to purple]. At lower concentrations, $\phi \leq 0.44$, $\sigma(\gamma)$ weakly peaks before dropping to the high-strain plateau. Such `stress overshoots' are observed in a variety of soft materials~\cite{Divoux2011StressOver,Koumakis2011Geltwostep,Koumakis2012HSGStartup}, though proposed mechanisms vary. These samples also exhibit weakly non-monotonic flow curves, indicating a degree of thixotropy and/or shear banding. At higher concentrations  $\phi>0.54$ we do not reach a stable, high-strain plateau (dotted lines). Instead the stress increases continuously and we observe visible sample fracture, \partfigref{fig:fc}{b}, so $\sigma(\dot\gamma)$ was not recorded.

\section{Discussion and Conclusions\label{sec:dis}}

We now compare the yield stresses obtained from these different protocols, exploring how they vary with $\phi$ and link to the strain-state of the suspension. Despite the complexity of the (time-dependent) micro-physical interactions in the cement paste, we find a striking resemblance to model adhesive non-Brownian suspensions.

\subsection{Frictional and frictionless yield stresses\label{sec:dis:sy}}

\begin{figure}
    \centering
    \includegraphics[width=\columnwidth]{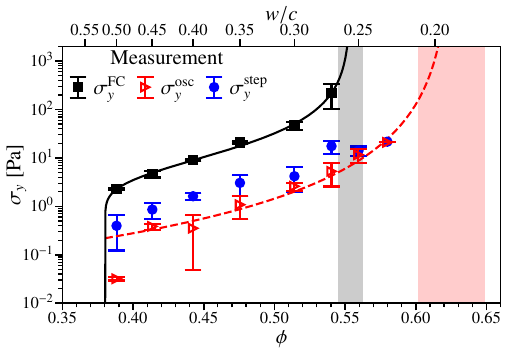}
    \caption{Compiled yield stresses with volume fraction (bottom axis) and $w/c$ (top). Flow curve stress down to an $w/c$ of 0.27 and oscillatory yield stresses down to 0.23. Solid line, fit to constraint model with $\phi_{\rm rlp} = 0.55(1)$ and $\phip = 0.38(1)$ with $\sigma_a = \SI{9(2)}{\pascal}$ and $\kappa = 0.8(2)$.  Dashed (red) line power law divergence of oscillatory yield stress, $\sigma_y^{\rm osc} = A (1-\phi/\phicp)^{-m}$ with $A = \SI{0.016}{\pascal}$, $\phicp = 0.63(2)$ and $m=2.8$ fixed. Lowest $\phi$ excluded from fit.}
    \label{fig:comp}
\end{figure}

Comparing the $\phi$-dependence of the three different yield stresses that we have measured \figref{fig:comp}, the flow curve yield stress $\sfc$ (black squares) stands apart. Across the full $\phi$ range $\sfc$ is significantly larger than the other two. Furthermore, at high concentrations $\phi\gtrsim 0.5$ we find that  $\sfc(\phi)$ grows increasingly rapidly and appears to diverge somewhere around 0.55 while $\sosc$ and $\sStep$ remain finite up to $\phi=0.58$.

While is has been noted that yield stress of cement pastes can depend on the measurement protocol~\cite{chougnet2007linear,yuan2017measurement}, the relation between protocol and the maximum `jamming' concentration has been largely unexplored. Here, motivated by similar work yielding in a model adhesive, non-Brownian suspension~\cite{richards2020role}, we first focus on comparing $\sfc$ and $\sosc$.

This prior work using cornstarch particles suspended in oil as a model adhesive, non-Brownian system found a similar discrepancy between the yield stress obtain from oscillatory tests and steady-shear flow curves, with the oscillatory yield stress diverging at a notably higher volume fraction. This reflects the frictional contact network underlying the steady-shear yield stress, so that $\sfc$ diverges at the frictional jamming point $\phim$, while the small amplitude oscillations disrupt these contacts so that $\sosc$ is instead controlled by a higher friction-less jamming limit $\approx\phicp$, the geometric random close packing limit. 

We can quantify these two distinct critical concentrations through fits to $\sfc(\phi)$ and $\sosc(\phi)$. Motivated by the qualitative similarities between our cement suspensions and the model system from \refcite{richards2020role}, we fit $\sfc(\phi)$ to a `constraint-based' model for the yielding of adhesive, frictional non-Brownian particles~\cite{guy2018constraint}. In this implementation, particles are frictional under all conditions, so that sliding motion is always constrained, but there is a stress scale $\sigma_a$ for peeling apart adhesive bonds to initiate rolling. The critical jamming concentration is thus stress-dependent, interpolating between a lower limit $\phip$, where contacting particles can neither slide nor roll, and an upper limit $\phim$ where sliding remains constrained but particles can now roll. 

The yield stress in this implementation of the model takes the form expressed in \refcite{larsen2023rheology},
\begin{equation}
    \sigma^{\rm FC}_y = \sigma_a \left[ \ln \left (\frac{\phim-\phip}{\phi-\phip} \right)\right]^{-1/\kappa},
\end{equation}
with a finite yield stress emerging at $\phip$, increasing with $\phi$ and ultimately diverging at $\phim$. Fitting $\sfc(\phi)$ to this expression gives $\sigma_a = \SI{9(2)}{\pascal}$ and $\kappa = 0.8(2)$, describing the release of adhesive bonds under stress~\cite{richards2020role}, and the critical volume fractions $\phip =  0.38(1)$ and $\phim = 0.55(1)$. While ``Yodel'' (yield stress model for suspensions) would provide a similar functional form~\cite{flatt2006yodel}, this constraint-based framework provided a novel physical interpretation of the critical volume fractions. Sedimentation prevents accurate measurements at lower volume fractions, $\phi <0.38$, suggesting an upper bound set by the gravitational stress scale $\Delta \rho g d \approx \SI{0.2}{\pascal}$ for $d \sim \SI{10}{\micro\metre}$, \partfigref{fig:psd}{b}, suggesting the rapid drop approaching $\phip$ is not unreasonable. 

In contrast to the divergence of $\sfc$ approaching $\phim\approx 0.55$, $\sosc$ remains finite for $\phi \leq 0.58$, showing only gradual steepening. We would expect that this yield stress should diverge around the close packing limit, $\phicp<1$, and thus use a purely empirical Krieger-Dougherty-like form~\cite{richards2020role} $\sosc = A (1-\phi/\phicp)^{-m}$ to estimate this upper critical concentration. We restrict this fit to $\phi>0.4$, as we expect this form to only apply approaching the divergence and away from any lower percolation limit. Fixing $m=2.8$ we find $\phicp = 0.63(2)$ with $A = \SI{0.016}{\pascal}$, \figref{fig:comp}.

While measurements of $\sosc$ at even higher $\phi$ could better constrain $\phicp$ and allow a robust estimate of $m$, higher-concentration samples either fractured or displayed notable drying during rheometer loading. We highlight the larger uncertainty in $\phicp$ versus $\phim$ (cf.~grey and red bands in \figref{fig:comp}), though there is still a significant gap between our upper bound on $\phim$ and our lower bound on $\phicp$. This gap, along with the observation that samples above $\phi=0.55$ fracture at large strains without reaching a steady-state, demonstrates that two distinct critical concentrations control $\sfc$ and $\sosc$. 

Packing tests with the Rigden apparatus suggest a maximum packing fraction of 0.68 for the dry cement powder, above the random close packing limit for uniform spheres ($\approx 0.64$), reflecting the high particle polydispersity~\cite{baranau2014random}. This dry packing limit is above our estimate of $\phicp\approx 0.63$ from the divergence of $\sosc$. However, we should expect changes to the particles in solution, with rapid processes such as the dissolution and recrystallisation of sulphate-containing phases~\cite{ylmen2009early} likely occurring within our mixing time. Thus, we can only consider this dry packing limit as an upper bound, so our results are consistent with $\sosc$ being limited by a near random close packing limit at $\phi \approx 0.63$.

This implies that $\sfc$ is controlled by a lower, frictional, jamming point, $\phim$, in contrast to previous models that take the critical point to be at maximum packing~\cite{flatt2006yodel}. In a frictional materials loads are transmitted via compressional ``force chains'', which are disrupted by changes in direction and require strain to reform~\cite{cates1998jamming}. The stress evolution during step-shear tests supports this frictional force-chain picture.

\subsection{Connection via step shear\label{sec:dis:step}}

The step shear test links an isotropic state, prepared by oscillatory shear with decreasing strain amplitude, to an anisotropic state, in the continuous shear flow curve. The step yield stress at $\gamma \approx 0.001$, $\sStep$, is then equivalent to $\sosc$, as small, initially elastic, displacements from an isotropic state. Comparing values, we see reasonable agreement, \cf\ (blue) circles and open (red) triangles, except at the lowest $\phi < 0.40$ where sedimentation becomes problematic. Further reinforcing this connection, both the oscillatory yield strain \partfigref{fig:osc}{b} and the this step yield strain \cf\ (blue) shading in \figref{fig:step} are similar in magnitude, $\gamma \sim 10^{-3}$, and independent of $\phi$.

Although close compared to $\sfc$, $\sStep$ is consistently 2--4$\times$ higher than $\sosc$ for $0.40 < \phi < 0.55$, beyond our estimated errors. This may be attributable to the higher shear rate for in the step strain tests, $\dot\gamma = \SI{0.05}{\per\second}$, compared to $\dot\gamma = \gamma_0\omega \approx \SI{0.005}{\per\second}$ for $\sosc$. This may also reflect our definition of $\sStep$ from the start of the stress plateau, as opposed to earlier deviations from a linear response. However, limitations set by the controlled-strain mode of our rheometer and sample drying at low rates (thus, longer test times) prevented further investigation.

After this initial yielding at small strains, the cement suspensions strain harden as  $\gamma$ increasing, with $\sigma(\gamma)$ increasing up to the steady shear yield stress $\sfc$ for $\phi<\phim$, \cf\ solid and dashed lines in \figref{fig:step}. This reflects the development of frictional contacts and force chains. At $\phi>\phim$, above the divergence of $\sfc$ [dotted (purple) lines], $\sigma(\gamma)$ continually increases, with fracture observed, and a steady, homogeneously flowing state is not reached. This qualitative change is seen with other suspensions above jamming, such as shear-thickening suspensions at high stress~\cite{richards2019competing,ovarlez2020density}, although details are system specific.

Defining a characteristic strain scale $\gamma^*$ for this hardening by $\sigma(\gamma^*) = 0.5\sigma_{\max}$~\cite{dambrosio2023role}, we can quantitatively compare the response of our cement pastes to other non-Brownian suspensions. We find $\gamma^*\approx 0.16$, largely independent of $\phi$, roughly corresponding to the `shoulder' in $\elastic(0.1< \gamma_0 < 1)$ see in oscillatory measurements, \partfigref{fig:osc}{a}. Experiments with non-adhesive suspensions at equivalent $\phi$ found that larger strains $\gamma \approx 1$ were needed to recover the flowing viscosity in shear reversal tests~\cite{dambrosio2023role,peters2016rheology}. This should be analogous to a strain of $\approx 0.5$ when starting from an isotropic state (instead of reversing between opposing anisotropic states), still notably larger than our measured $\gamma^*$. Furthermore, their reversal strain grows with decreasing $\phi$, suggesting the detailed dynamics depend on the particle interactions and the nature of the frictional contacts, potentially involving aggregated structures.

\subsection{Outlook\label{sec:dis:out}}

By comparing multiple yield stress measurement techniques on a fresh Portland cement suspension over a wide range of $w/c$ ratios (and so volume fractions), we reveal a  strong dependence on preparation protocol. In particular, the yield stress for a cement suspension prepared in an anisotropic state by continuous shear, $\sfc$, is larger and diverges at a lower volume fraction, $\phim$, compared to the yield stress for an isotropic state prepared by oscillatory shear, $\sosc$. We associate the larger $\sfc$ with the development of a frictional contact network and the smaller $\sosc$ with the underlying gel-like attractive nature of the particles. The strain-dependent change between these two values is explicitly shown by a step shear test. Together, this suggests that the yield stress of a conventional Portland-cement--based suspension, mortar or concrete for, \eg\, pumping down a pipe or for extrusion in additive manufacture, is strongly influenced by frictional forces between cement particles. The yield stress is therefore tied to the frictional jamming point, which can be substantially below the frictionless random close packing limit.

The link to $\phim$ suggests that the rheology of a fresh cement suspension can be mapped to a non-setting suspension, such as calcite~\cite{richards2021turning}, from which the effects of setting can be considered. Such pastes have been previously developed as a potential reference material~\cite{ferraris2014development} and used for studying admixtures~\cite{sha2023superplasticizers}. 

Linking $\lesssim \SI{100}{\micro\metre}$ cement particles, not just $\gtrsim \SI{1}{\milli\metre}$ aggregates, to frictional packing opens up a host of further questions. As $\sfc$ reflects an anisotropic sheared state, the suspension is sensitive to changes in load direction. Combined with advances in controlling jammed suspensions by alternating shear direction~\cite{ness2018shaken,acharya2023optimum}, the vibrational compaction of concrete may be optimised based on the cement background. However, the set state this is clearly not anisotropic. Simulations of setting and its influence on structure~\cite{ioannidou2016crucial} may shed light on this transition. 

Ultimately, the method and protocols detailed in this work provide scope for the modification of Portland cement with an understanding at the particle level using the separation of frictional, granular-like properties and attractive colloidal-like contributions. Of particular interest may be the effect of cement substitution with, \eg\ silica~\cite{wu2019changes}, fly-ash~\cite{jiang2020utilization} or blast furnace slag~\cite{ting2019effects}. Further, the role of admixitures such as super-plasticisers for self-compacting concrete could be rationalised and optimised with an understanding of friction modification vs reducing attraction.

In both cases, for a full understanding the setting reaction may also need to be considered, as this controls surface roughness~\cite{ioannidou2016mesoscale} that is essential for inter-particle friction. This highlights the multi-scale nature of cement-based materials, with understanding needed from the chemical nanoscale to microscale particle surface properties, then to local flow properties, and finally to inhomogeneous bulk flows containing aggregates, for which this work provides a mesoscale link.

\section*{Acknowledgements}

The authors thank Rebecca Rae for X-ray powder diffraction, and Chris Ness, Wilson Poon, Job Thijssen and Daniel Hodgson for discussions. The authors acknowledge access to the Cryo-FIB-SEM Imaging Facility for Opaque Soft Matter and the CSEC-Based Facility for X-ray Characterisation of Materials (UK Engineering and Physical Sciences Research Council grant nos EP/P030564/1 and EP/V03605X/1). The datasets generated during and/or analysed during the current study will be available in the Edinburgh DataShare repository, \url{https://doi.org/10.7488/ds/XXXX}. CRediT authorship contribution statement. \textbf{J.A.R.}: Conceptualisation, Formal analysis, Investigation (supporting), Supervision (supporting), Writing - original draft, Writing - review \& editing. \textbf{H.L.}: Formal analysis, Investigation (lead), Validation. \textbf{R.E.O'N.}: Conceptualisation, Resources, Supervision (supporting). \textbf{F.H.J.L.}: Investigation (SEM). \textbf{J.R.R.}: Conceptualisation, Supervision (lead), Writing - review \& editing.

\appendix
\setcounter{figure}{0}

\section{Sample details}

See Table~\ref{tab:repeats} for number of included repeats.

\begin{table}
    \centering
    \begin{tabularx}{\columnwidth}{XXX}
        \hline
         Water-to-cement ratio, $w/c$\vspace{0.2em}&  Volume fraction, $\phi$ & No.\ of repeats\\
         \hline
         0.50&  0.39& 3 \\
         0.45&  0.41& 4\\
         0.40&  0.44& 3\\
         0.35&  0.48& 3\\
         0.30&  0.51& 5\\
         0.27&  0.54& 3\\
         0.25&  0.56& 2\\
         0.23&  0.58& 1\\
         0.21&  0.60& $-$\\
         \hline \vspace{0.1em}
    \end{tabularx}
    \caption{Sample details and number of repeat measurements. The variability in the number of repeats for $w/c\geq0.27$ reflects discarded samples due to loading errors with significant squeeze flow, and no samples for $w/c \geq 0.21$ the inability to load due to paste stiffness and fracture.}
    \label{tab:repeats}
\end{table}

\section{Decreasing strain amplitude sweeps}\label{app:prep}

\begin{figure*}
    \centering
    \includegraphics[width=\textwidth]{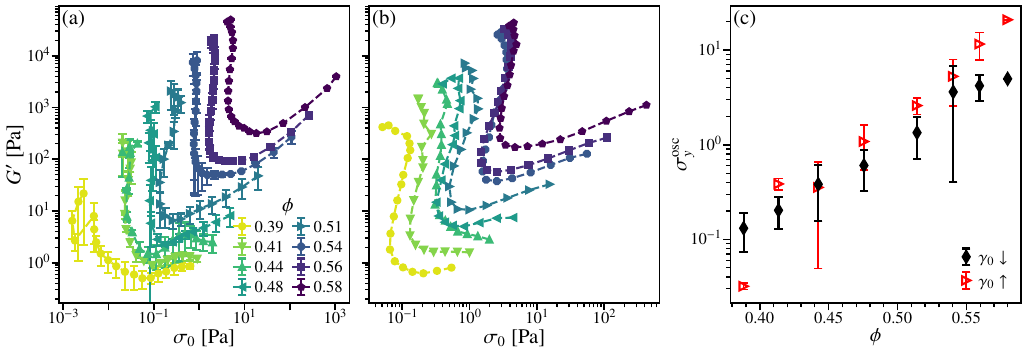}
    \caption{Oscillatory preparation protocol with decreasing strain amplitude sweep. (a) Preparation from freshly loaded state with decreasing strain amplitude sweep. Symbols, elastic modulus ($G^{\prime}$) with oscillatory stress, $\sigma_0$, dashed lines to guide the eye. Increasing solid volume fraction from light (yellow) to dark (blue), see inset legend. (b)~Corresponding preparation step prior to continuous shear. (c)~Comparison of oscillatory yield stress measurements over varying volume fraction. Symbols: oscillatory yield stress, $\sosc$, extracted from decreasing strain amplitude sweep in (b), (black) diamonds, \textit{vs} increasing strain amplitude sweep in \partfigref{fig:osc}{a}, (red) triangles.}
    \label{fig:oscDown}
\end{figure*}

During preparation, an oscillatory strain amplitude sweep is applied from $\gamma_0 = 0.3$ to $10^{-4}$. Plotting $\elastic(\sigma_0)$ directly after loading, \partfigref{fig:oscDown}{a}, and after the oscillatory test, \partfigref{fig:oscDown}{b}, the yield point appears well-defined, as $\elastic(\sigma_0)$ becomes vertical or even re-entrant. A yield stress can be defined from the `nose' in $\elastic(\sigma_0)$, as for a non-setting suspension~\cite{richards2020role}, $\loss$ is not shown in the full range of $\gamma_0$ for clarity. $\sosc$ for $\gamma_0 \downarrow$, varies from \SIrange{0.1}{5}{\pascal} for $\phi = 0.39$ to 0.58, light (yellow) to dark (blue), \partfigref{fig:oscDown}{c}. At the lowest $\phi = 0.39$ (yellow), the measured stress, $\SI{0.1}{\pascal}$, is higher than the preceding $\sosc = \SI{0.03}{\pascal}$. In contrast, at $\phi \geq 0.54$, the measured stress remains $\approx \SI{5}{\pascal}$, while $\sosc$ increases from \SIrange{5}{24}{\pascal}. However, across a broad intermediate range of $\phi$, they are within a factor $\approx 2$. The origin in the change in form between increasing and decreasing amplitude sweeps remains an area for future investigation, in particular disentangling the role of contact evolution over time~\cite{bonacci2020contact} vs structural evolution with strain~\cite{clarke2021gel}.


\begin{thebibliography}{10}
\expandafter\ifx\csname url\endcsname\relax
  \def\url#1{\texttt{#1}}\fi
\expandafter\ifx\csname urlprefix\endcsname\relax\def\urlprefix{URL }\fi
\expandafter\ifx\csname href\endcsname\relax
  \def\href#1#2{#2} \def\path#1{#1}\fi

\bibitem{miller2020climate}
S.~A. Miller, F.~C. Moore, Climate and health damages from global concrete production, Nat. Clim. Chang. 10~(5) (2020) 439--443.
\newblock \href {https://doi.org/10.1038/s41558-020-0733-0} {\path{doi:10.1038/s41558-020-0733-0}}.

\bibitem{gagg2014cement}
C.~R. Gagg, Cement and concrete as an engineering material: {A}n historic appraisal and case study analysis, Eng. Fail. Anal. 40 (2014) 114--140.
\newblock \href {https://doi.org/10.1016/j.engfailanal.2014.02.004} {\path{doi:10.1016/j.engfailanal.2014.02.004}}.

\bibitem{li2020durability}
J.~Li, Z.~Wu, C.~Shi, Q.~Yuan, Z.~Zhang, Durability of ultra-high performance concrete--{A} review, Constr. Build. Mater. 255 (2020) 119296.
\newblock \href {https://doi.org/10.1016/j.conbuildmat.2020.119296} {\path{doi:10.1016/j.conbuildmat.2020.119296}}.

\bibitem{bentz1999effects}
D.~P. Bentz, E.~J. Garboczi, C.~J. Haecker, O.~M. Jensen, Effects of cement particle size distribution on performance properties of {P}ortland cement-based materials, Cem. Concr. Res. 29~(10) (1999) 1663--1671.
\newblock \href {https://doi.org/10.1016/S0008-8846(99)00163-5} {\path{doi:10.1016/S0008-8846(99)00163-5}}.

\bibitem{ioannidou2016crucial}
K.~Ioannidou, M.~Kandu{\v{c}}, L.~Li, D.~Frenkel, J.~Dobnikar, E.~Del~Gado, The crucial effect of early-stage gelation on the mechanical properties of cement hydrates, Nat. Commun. 7~(1) (2016) 12106.
\newblock \href {https://doi.org/10.1038/ncomms12106} {\path{doi:10.1038/ncomms12106}}.

\bibitem{zhang2011tests}
Appendix - tests of building materials, in: H.~Zhang (Ed.), Building Materials in Civil Engineering, Woodhead Publishing Series in Civil and Structural Engineering, Woodhead Publishing, 2011, pp. 344--422.
\newblock \href {https://doi.org/10.1533/9781845699567.appendix} {\path{doi:10.1533/9781845699567.appendix}}.

\bibitem{yammine2008ordinary}
J.~Yammine, M.~Chaouche, M.~Guerinet, M.~Moranville, N.~Roussel, From ordinary rhelogy concrete to self compacting concrete: {A} transition between frictional and hydrodynamic interactions, Cem. Conc. Res. 38~(7) (2008) 890--896.
\newblock \href {https://doi.org/j.cemconres.2008.03.011} {\path{doi:j.cemconres.2008.03.011}}.

\bibitem{choi2013lubrication}
M.~Choi, N.~Roussel, Y.~Kim, J.~Kim, Lubrication layer properties during concrete pumping, Cem. Conc. Res. 45 (2013) 69--78.
\newblock \href {https://doi.org/j.cemconres.2012.11.001} {\path{doi:j.cemconres.2012.11.001}}.

\bibitem{saak1999characterization}
A.~W. Saak, H.~M. Jennings, S.~P. Shah, Characterization of the rheological properties of cement paste for use in self-compacting concrete, in: {\^A}.~Skarendahl, {\"O}.~Petersson (Eds.), Proceedings of the 1st International RILEM Symposium on Self-Compacting Concrete, RILEM Publications SARL, Bagneux, France, 1999, pp. 83--93.

\bibitem{ovarlez2006physical}
G.~Ovarlez, N.~Roussel, A physical model for the prediction of lateral stress exerted by self-compacting concrete on formwork, Mater. Struct. 39 (2006) 269--279.
\newblock \href {https://doi.org/10.1617/s11527-005-9052-1} {\path{doi:10.1617/s11527-005-9052-1}}.

\bibitem{roussel2018rheological}
N.~Roussel, Rheological requirements for printable concretes, Cem. Conc. Res. 112 (2018) 76--85.
\newblock \href {https://doi.org/j.cemconres.2018.04.005} {\path{doi:j.cemconres.2018.04.005}}.

\bibitem{worrell2001carbon}
E.~Worrell, L.~Price, N.~Martin, C.~Hendriks, L.~O. Meida, Carbon dioxide emissions from the global cement industry, Annu. Rev. Environ. Resour. 26~(1) (2001) 303--329.
\newblock \href {https://doi.org/10.1146/annurev.energy.26.1.303} {\path{doi:10.1146/annurev.energy.26.1.303}}.

\bibitem{panesar2020performance}
D.~K. Panesar, R.~Zhang, Performance comparison of cement replacing materials in concrete: {L}imestone fillers and supplementary cementing materials--a review, Constr. Build. Mater. 251 (2020) 118866.
\newblock \href {https://doi.org/10.1016/j.conbuildmat.2020.118866} {\path{doi:10.1016/j.conbuildmat.2020.118866}}.

\bibitem{ferraris2001influence}
C.~F. Ferraris, K.~H. Obla, R.~Hill, The influence of mineral admixtures on the rheology of cement paste and concrete, Cem. Concr. Res. 31~(2) (2001) 245--255.
\newblock \href {https://doi.org/10.1016/S0008-8846(00)00454-3} {\path{doi:10.1016/S0008-8846(00)00454-3}}.

\bibitem{jiang2020utilization}
D.~Jiang, X.~Li, Y.~Lv, M.~Zhou, C.~He, W.~Jiang, Z.~Liu, C.~Li, Utilization of limestone powder and fly ash in blended cement: {R}heology, strength and hydration characteristics, Constr. Build. Mater. 232 (2020) 117228.
\newblock \href {https://doi.org/10.1016/j.conbuildmat.2019.117228} {\path{doi:10.1016/j.conbuildmat.2019.117228}}.

\bibitem{wu2019changes}
Z.~Wu, K.~H. Khayat, C.~Shi, Changes in rheology and mechanical properties of ultra-high performance concrete with silica fume content, Cem. Concr. Res. 123 (2019) 105786.
\newblock \href {https://doi.org/10.1016/j.cemconres.2019.105786} {\path{doi:10.1016/j.cemconres.2019.105786}}.

\bibitem{ting2019effects}
L.~Ting, W.~Qiang, Z.~Shiyu, Effects of ultra-fine ground granulated blast-furnace slag on initial setting time, fluidity and rheological properties of cement pastes, Powder Technol. 345 (2019) 54--63.
\newblock \href {https://doi.org/10.1016/j.powtec.2018.12.094} {\path{doi:10.1016/j.powtec.2018.12.094}}.

\bibitem{boyer2011unifying}
F.~Boyer, {\'E}.~Guazzelli, O.~Pouliquen, Unifying suspension and granular rheology, Phys. Rev. Lett. 107~(18) (2011) 188301.
\newblock \href {https://doi.org/10.1103/PhysRevLett.107.188301} {\path{doi:10.1103/PhysRevLett.107.188301}}.

\bibitem{guazzelli2018rheology}
{\'E}.~Guazzelli, O.~Pouliquen, Rheology of dense granular suspensions, J. Fluid Mech. 852 (2018) P1.
\newblock \href {https://doi.org/10.1017/jfm.2018.548} {\path{doi:10.1017/jfm.2018.548}}.

\bibitem{royall2013search}
C.~P. Royall, W.~C.~K. Poon, E.~R. Weeks, In search of colloidal hard spheres, Soft Matter 9~(1) (2013) 17--27.
\newblock \href {https://doi.org/10.1039/C2SM26245B} {\path{doi:10.1039/C2SM26245B}}.

\bibitem{silbert2010jamming}
L.~E. Silbert, Jamming of frictional spheres and random loose packing, Soft Matter 6~(13) (2010) 2918--2924.
\newblock \href {https://doi.org/10.1039/C001973A} {\path{doi:10.1039/C001973A}}.

\bibitem{guy2015towards}
B.~M. Guy, M.~Hermes, W.~C.~K. Poon, Towards a unified description of the rheology of hard-particle suspensions, Phys. Rev. Lett. 115~(8) (2015) 088304.
\newblock \href {https://doi.org/10.1103/PhysRevLett.115.088304} {\path{doi:10.1103/PhysRevLett.115.088304}}.

\bibitem{clavaud2017revealing}
C.~Clavaud, A.~B{\'e}rut, B.~Metzger, Y.~Forterre, Revealing the frictional transition in shear-thickening suspensions, Proc. Natl. Acad. Sci. U.S.A. 114~(20) (2017) 5147--5152.
\newblock \href {https://doi.org/10.1073/pnas.1703926114} {\path{doi:10.1073/pnas.1703926114}}.

\bibitem{lin2015hydrodynamic}
N.~Y.~C. Lin, B.~M. Guy, M.~Hermes, C.~Ness, J.~Sun, W.~C.~K. Poon, I.~Cohen, Hydrodynamic and contact contributions to continuous shear thickening in colloidal suspensions, Phys. Rev. Lett. 115~(22) (2015) 228304.
\newblock \href {https://doi.org/10.1103/PhysRevLett.115.228304} {\path{doi:10.1103/PhysRevLett.115.228304}}.

\bibitem{comtet2017pairwise}
J.~Comtet, G.~Chatt{\'e}, A.~Nigues, L.~Bocquet, A.~Siria, A.~Colin, Pairwise frictional profile between particles determines discontinuous shear thickening transition in non-colloidal suspensions, Nat. Commun. 8~(1) (2017) 15633.
\newblock \href {https://doi.org/10.1038/ncomms15633} {\path{doi:10.1038/ncomms15633}}.

\bibitem{wyart2014discontinuous}
M.~Wyart, M.~E. Cates, Discontinuous shear thickening without inertia in dense non-{B}rownian suspensions, Phys. Rev. Lett. 112~(9) (2014) 098302.
\newblock \href {https://doi.org/10.1103/PhysRevLett.112.098302} {\path{doi:10.1103/PhysRevLett.112.098302}}.

\bibitem{guy2018constraint}
B.~M. Guy, J.~A. Richards, D.~J.~M. Hodgson, E.~Blanco, W.~C.~K. Poon, Constraint-based approach to granular dispersion rheology, Phys. Rev. Lett. 121~(12) (2018) 128001.
\newblock \href {https://doi.org/10.1103/PhysRevLett.121.128001} {\path{doi:10.1103/PhysRevLett.121.128001}}.

\bibitem{richards2020role}
J.~A. Richards, B.~M. Guy, E.~Blanco, M.~Hermes, G.~Poy, W.~C.~K. Poon, The role of friction in the yielding of adhesive non-{B}rownian suspensions, J. Rheol. 64~(2) (2020) 405--412.
\newblock \href {https://doi.org/10.1122/1.5132395} {\path{doi:10.1122/1.5132395}}.

\bibitem{richards2019competing}
J.~A. Richards, J.~R. Royer, B.~Liebchen, B.~M. Guy, W.~C.~K. Poon, Competing timescales lead to oscillations in shear-thickening suspensions, Phys. Rev. Lett. 123~(3) (2019) 038004.
\newblock \href {https://doi.org/10.1103/PhysRevLett.123.038004} {\path{doi:10.1103/PhysRevLett.123.038004}}.

\bibitem{han2019stress}
E.~Han, N.~M. James, H.~M. Jaeger, Stress controlled rheology of dense suspensions using transient flows, Phys. Rev. Lett. 123~(24) (2019) 248002.
\newblock \href {https://doi.org/10.1103/PhysRevLett.123.248002} {\path{doi:10.1103/PhysRevLett.123.248002}}.

\bibitem{stutzman2016phase}
P.~E. Stutzman, P.~Feng, J.~W. Bullard, Phase analysis of {P}ortland cement by combined quantitative {X}-ray powder diffraction and scanning electron microscopy, J. Res. Natl. Inst. Stand. Technol. 121 (2016) 47.
\newblock \href {https://doi.org/10.6028/jres.121.004} {\path{doi:10.6028/jres.121.004}}.

\bibitem{hackley2004particle}
V.~A. Hackley, V.~Gintautas, C.~F. Ferraris, Particle size analysis by laser diffraction spectrometry: {A}pplication to cementitious powders, US Department of Commerce, National Institute of Standards and Technology, Gaithersburg, MD (USA), 2004.
\newblock \href {https://doi.org/10.6028/NIST.IR.7097} {\path{doi:10.6028/NIST.IR.7097}}.

\bibitem{song2021occurrence}
Q.~Song, J.~Su, J.~Nie, H.~Li, Y.~Hu, Y.~Chen, R.~Li, Y.~Deng, The occurrence of MgO and its influence on properties of clinker and cement: A review, Constr. Build. Mater. 293 (2021) 123494.
\newblock \href {https://doi.org/10.1016/j.conbuildmat.2021.123494} {\path{doi:10.1016/j.conbuildmat.2021.123494}}.

\bibitem{flatt2004dispersion}
R.~J. Flatt, Dispersion forces in cement suspensions, Cem. Concr. Res. 34~(3) (2004) 399--408.
\newblock \href {https://doi.org/10.1016/j.cemconres.2003.08.019} {\path{doi:10.1016/j.cemconres.2003.08.019}}.

\bibitem{rigden1947use}
P.~J. Rigden, The use of fillers in bituminous road surfacings. a study of filler-binder systems in relation to filler characteristics, J. Soc. Chem. Ind. 66~(9) (1947) 299--309.
\newblock \href {https://doi.org/10.1002/jctb.5000660902} {\path{doi:10.1002/jctb.5000660902}}.

\bibitem{vance2015rheology}
K.~Vance, G.~Sant, N.~Neithalath, The rheology of cementitious suspensions: a closer look at experimental parameters and property determination using common rheological models, Cem. Concr. Compos. 59 (2015) 38--48.
\newblock \href {https://doi.org/10.1016/j.cemconcomp.2015.03.001} {\path{doi:10.1016/j.cemconcomp.2015.03.001}}.

\bibitem{delhaye2000squeeze}
N.~Delhaye, A.~Poitou, M.~Chaouche, Squeeze flow of highly concentrated suspensions of spheres, J. Non-Newton. Fluid Mech. 94~(1) (2000) 67--74.
\newblock \href {https://doi.org/10.1016/S0377-0257(00)00130-0} {\path{doi:10.1016/S0377-0257(00)00130-0}}.

\bibitem{grandes2021rheological}
F.~A. Grandes, V.~K. Sakano, A.~C. Rego, M.~S. Rebmann, F.~A. Cardoso, R.~G. Pileggi, Rheological behavior and flow induced microstructural changes of cement-based mortars assessed by pressure mapped squeeze flow, Powder Technology 393 (2021) 519--538.
\newblock \href {https://doi.org/10.1016/j.powtec.2021.07.082} {\path{doi:10.1016/j.powtec.2021.07.082}}.

\bibitem{raghavan1995shear}
S.~R. Raghavan, S.~A. Khan, Shear-induced microstructural changes in flocculated suspensions of fumed silica, J. Rheol. 39~(6) (1995) 1311--1325.
\newblock \href {https://doi.org/10.1122/1.550638} {\path{doi:10.1122/1.550638}}.

\bibitem{lecampion2014confined}
B.~Lecampion, D.~I. Garagash, Confined flow of suspensions modelled by a frictional rheology, J. Fluid Mech. 759 (2014) 197--235.
\newblock \href {https://doi.org/10.1017/jfm.2014.557} {\path{doi:10.1017/jfm.2014.557}}.

\bibitem{fall2010shear}
A.~Fall, A.~Lemaitre, F.~Bertrand, D.~Bonn, G.~Ovarlez, Shear thickening and migration in granular suspensions, Phys. Rev. Lett. 105~(26) (2010) 268303.
\newblock \href {https://doi.org/10.1103/PhysRevLett.105.268303} {\path{doi:10.1103/PhysRevLett.105.268303}}.

\bibitem{carotenuto2015predicting}
C.~Carotenuto, A.~Vananroye, J.~Vermant, M.~Minale, Predicting the apparent wall slip when using roughened geometries: {A} porous medium approach, J. Rheol. 59~(5) (2015) 1131--1149.
\newblock \href {https://doi.org/10.1122/1.4923405} {\path{doi:10.1122/1.4923405}}.

\bibitem{richards2021turning}
J.~A. Richards, R.~E. O’Neill, W.~C.~K. Poon, Turning a yield-stress calcite suspension into a shear-thickening one by tuning inter-particle friction, Rheol. Acta 60 (2021) 97--106.
\newblock \href {https://doi.org/10.1007/s00397-020-01247-z} {\path{doi:10.1007/s00397-020-01247-z}}.

\bibitem{ovarlez2013existence}
G.~Ovarlez, S.~Cohen-Addad, K.~Krishan, J.~Goyon, P.~Coussot, On the existence of a simple yield stress fluid behavior, J. Non-Newton Fluid Mech. 193 (2013) 68--79.
\newblock \href {https://doi.org/10.1016/j.jnnfm.2012.06.009} {\path{doi:10.1016/j.jnnfm.2012.06.009}}.

\bibitem{meeker2004slip_a}
S.~P. Meeker, R.~T. Bonnecaze, M.~Cloitre, Slip and flow in soft particle pastes, Phys. Rev. Lett. 92~(19) (2004) 198302.
\newblock \href {https://doi.org/10.1103/PhysRevLett.92.198302} {\path{doi:10.1103/PhysRevLett.92.198302}}.

\bibitem{meeker2004slip_b}
S.~P. Meeker, R.~T. Bonnecaze, M.~Cloitre, Slip and flow in pastes of soft particles: {D}irect observation and rheology, J. Rheol. 48~(6) (2004) 1295--1320.
\newblock \href {https://doi.org/10.1122/1.1795171} {\path{doi:10.1122/1.1795171}}.

\bibitem{owens2020improved}
C.~E. Owens, A.~J. Hart, G.~H. McKinley, Improved rheometry of yield stress fluids using bespoke fractal {3D} printed vanes, J. Rheol. 64~(3) (2020) 643--662.
\newblock \href {https://doi.org/10.1122/1.5132340} {\path{doi:10.1122/1.5132340}}.

\bibitem{chaparian2022computational}
E.~Chaparian, C.~E. Owens, G.~H. McKinley, Computational rheometry of yielding and viscoplastic flow in vane-and-cup rheometer fixtures, J. Non-Newton Fluid Mech. 307 (2022) 104857.
\newblock \href {https://doi.org/10.1016/j.jnnfm.2022.104857} {\path{doi:10.1016/j.jnnfm.2022.104857}}.

\bibitem{liberto2022small}
T.~Liberto, M.~Bellotto, A.~Robisson, Small oscillatory rheology and cementitious particle interactions, Cem. Concr. Res. 157 (2022) 106790.
\newblock \href {https://doi.org/10.1016/j.cemconres.2022.106790} {\path{doi:10.1016/j.cemconres.2022.106790}}.

\bibitem{bellotto2013cement}
M.~Bellotto, Cement paste prior to setting: {A} rheological approach, Cem. Concr. Res. 52 (2013) 161--168.
\newblock \href {https://doi.org/10.1016/j.cemconres.2013.07.002} {\path{doi:10.1016/j.cemconres.2013.07.002}}.

\bibitem{nachbaur2001dynamic}
L.~Nachbaur, J.~C. Mutin, A.~Nonat, L.~Choplin, Dynamic mode rheology of cement and tricalcium silicate pastes from mixing to setting, Cem. Concr. Res. 31~(2) (2001) 183--192.
\newblock \href {https://doi.org/10.1016/S0008-8846(00)00464-6} {\path{doi:10.1016/S0008-8846(00)00464-6}}.

\bibitem{dinkgreve2016different}
M.~Dinkgreve, J.~Paredes, M.~M. Denn, D.~Bonn, On different ways of measuring ``the''' yield stress, J. Non-Newton Fluid Mech. 238 (2016) 233--241.
\newblock \href {https://doi.org/10.1016/j.jnnfm.2016.11.001} {\path{doi:10.1016/j.jnnfm.2016.11.001}}.

\bibitem{richards2021characterising}
J.~A. Richards, V.~A. Martinez, J.~Arlt, Characterising shear-induced dynamics in flowing complex fluids using differential dynamic microscopy, Soft Matter 17~(39) (2021) 8838--8849.
\newblock \href {https://doi.org/10.1039/D1SM01094H} {\path{doi:10.1039/D1SM01094H}}.

\bibitem{hyun2011review}
K.~Hyun, M.~Wilhelm, C.~O. Klein, K.~S. Cho, J.~G. Nam, K.~H. Ahn, S.~J. Lee, R.~H. Ewoldt, G.~H. McKinley, A review of nonlinear oscillatory shear tests: Analysis and application of large amplitude oscillatory shear ({LAOS}), Progress in Polymer Science 36~(12) (2011) 1697--1753.
\newblock \href {https://doi.org/10.1016/j.progpolymsci.2011.02.002} {\path{doi:10.1016/j.progpolymsci.2011.02.002}}.

\bibitem{dinkgreve2017everything}
M.~Dinkgreve, M.~M. Denn, D.~Bonn, “everything flows?”: elastic effects on startup flows of yield-stress fluids, Rheol. Acta 56 (2017) 189--194.
\newblock \href {https://doi.org/10.1007/s00397-017-0998-z} {\path{doi:10.1007/s00397-017-0998-z}}.

\bibitem{Divoux2011StressOver}
T.~Divoux, C.~Barentin, S.~Manneville, Stress overshoot in a simple yield stress fluid: An extensive study combining rheology and velocimetry, Soft Matter 7 (2011) 9335--9349.
\newblock \href {https://doi.org/10.1039/C1SM05740E} {\path{doi:10.1039/C1SM05740E}}.

\bibitem{Koumakis2011Geltwostep}
N.~Koumakis, G.~Petekidis, Two step yielding in attractive colloids: transition from gels to attractive glasses, Soft Matter 7 (2011) 2456--2470.
\newblock \href {https://doi.org/10.1039/C0SM00957A} {\path{doi:10.1039/C0SM00957A}}.

\bibitem{Koumakis2012HSGStartup}
N.~Koumakis, M.~Laurati, S.~U. Egelhaaf, J.~F. Brady, G.~Petekidis, Yielding of hard-sphere glasses during start-up shear, Phys. Rev. Lett. 108 (2012) 098303.
\newblock \href {https://doi.org/10.1103/PhysRevLett.108.098303} {\path{doi:10.1103/PhysRevLett.108.098303}}.

\bibitem{chougnet2007linear}
A.~Chougnet, A.~Audibert, M.~Moan, Linear and non-linear rheological behaviour of cement and silica suspensions. {E}ffect of polymer addition, Rheol. Acta 46 (2007) 793--802.
\newblock \href {https://doi.org/10.1007/s00397-006-0126-y} {\path{doi:10.1007/s00397-006-0126-y}}.

\bibitem{yuan2017measurement}
Q.~Yuan, D.~Zhou, K.~H. Khayat, D.~Feys, C.~Shi, On the measurement of evolution of structural build-up of cement paste with time by static yield stress test vs. small amplitude oscillatory shear test, Cem. Concr. Res. 99 (2017) 183--189.
\newblock \href {https://doi.org/10.1016/j.cemconres.2017.05.014} {\path{doi:10.1016/j.cemconres.2017.05.014}}.

\bibitem{larsen2023rheology}
T.~Larsen, A.~S{\o}bye, J.~Royer, W.~Poon, T.~Larsen, S.~Andreasen, A.~Drozdov, J.~D.~C. Christiansen, Rheology of polydisperse nonspherical graphite particles suspended in mineral oil, J. Rheol. 67~(1) (2023) 81--89.
\newblock \href {https://doi.org/10.1122/8.0000511} {\path{doi:10.1122/8.0000511}}.

\bibitem{flatt2006yodel}
R.~J. Flatt, P.~Bowen, Yodel: a yield stress model for suspensions, J. Am. Ceram. Soc. 89~(4) (2006) 1244--1256.
\newblock \href {https://doi.org/10.1111/j.1551-2916.2005.00888.x} {\path{doi:10.1111/j.1551-2916.2005.00888.x}}.

\bibitem{baranau2014random}
V.~Baranau, U.~Tallarek, Random-close packing limits for monodisperse and polydisperse hard spheres, Soft Matter 10~(21) (2014) 3826--3841.
\newblock \href {https://doi.org/10.1039/C3SM52959B} {\path{doi:10.1039/C3SM52959B}}.

\bibitem{ylmen2009early}
R.~Ylm{\'e}n, U.~J{\"a}glid, B.-M. Steenari, I.~Panas, Early hydration and setting of {P}ortland cement monitored by {IR}, {SEM} and {V}icat techniques, Cem. Concr. Res. 39~(5) (2009) 433--439.
\newblock \href {https://doi.org/10.1016/j.cemconres.2009.01.017} {\path{doi:10.1016/j.cemconres.2009.01.017}}.

\bibitem{cates1998jamming}
M.~E. Cates, J.~P. Wittmer, J.-P. Bouchaud, P.~Claudin, Jamming, force chains, and fragile matter, Phys. Rev. Lett. 81~(9) (1998) 1841.
\newblock \href {https://doi.org/10.1103/PhysRevLett.81.1841} {\path{doi:10.1103/PhysRevLett.81.1841}}.

\bibitem{ovarlez2020density}
G.~Ovarlez, A.~Vu~Nguyen~Le, W.~J. Smit, A.~Fall, R.~Mari, G.~Chatt{\'e}, A.~Colin, Density waves in shear-thickening suspensions, Sci. Adv. 6~(16) (2020) eaay5589.
\newblock \href {https://doi.org/10.1126/sciadv.aay5589} {\path{doi:10.1126/sciadv.aay5589}}.

\bibitem{dambrosio2023role}
E.~d'Ambrosio, D.~L. Koch, S.~Hormozi, The role of rolling resistance in the rheology of wizarding quidditch ball suspensions, J. Fluid Mech. 974 (2023) A36.
\newblock \href {https://doi.org/10.1017/jfm.2023.756} {\path{doi:10.1017/jfm.2023.756}}.

\bibitem{peters2016rheology}
F.~Peters, G.~Ghigliotti, S.~Gallier, F.~Blanc, E.~Lemaire, L.~Lobry, Rheology of non-{B}ownian suspensions of rough frictional particles under shear reversal: {A} numerical study, J. Rheol. 60~(4) (2016) 715--732.
\newblock \href {https://doi.org/10.1122/1.4954250} {\path{doi:10.1122/1.4954250}}.

\bibitem{ferraris2014development}
C.~F. Ferraris, N.~S. Martys, W.~L. George, Development of standard reference materials for rheological measurements of cement-based materials, Cem. Concr. Compos. 54 (2014) 29--33.
\newblock \href {https://doi.org/10.1016/j.cemconcomp.2014.01.008} {\path{doi:10.1016/j.cemconcomp.2014.01.008}}.

\bibitem{sha2023superplasticizers}
S.~Sha, S.~Mantellato, S.~A. Weckwerth, Z.~Zhang, C.~Shi, R.~J. Flatt, Do superplasticizers work the way we think? {N}ew insights from their effect on the percolation threshold of limestone pastes, Cem. Concr. Res. 172 (2023) 107235.
\newblock \href {https://doi.org/10.1016/j.cemconres.2023.107235} {\path{doi:10.1016/j.cemconres.2023.107235}}.

\bibitem{ness2018shaken}
C.~Ness, R.~Mari, M.~E. Cates, Shaken and stirred: Random organization reduces viscosity and dissipation in granular suspensions, Sci. Adv. 4~(3) (2018) eaar3296.
\newblock \href {https://doi.org/10.1126/sciadv.aar3296} {\path{doi:10.1126/sciadv.aar3296}}.

\bibitem{acharya2023optimum}
P.~Acharya, M.~Trulsson, Optimum dissipation by cruising in dense suspensions (2023).
\newblock \href {http://arxiv.org/abs/2302.08810} {\path{arXiv:2302.08810}}.

\bibitem{ioannidou2016mesoscale}
K.~Ioannidou, K.~J. Krakowiak, M.~Bauchy, C.~G. Hoover, E.~Masoero, S.~Yip, F.-J. Ulm, P.~Levitz, R.~J.-M. Pellenq, E.~Del~Gado, Mesoscale texture of cement hydrates, Proc. Natl. Acad. Sci. U.S.A. 113~(8) (2016) 2029--2034.
\newblock \href {https://doi.org/10.1073/pnas.1520487113} {\path{doi:10.1073/pnas.1520487113}}.

\bibitem{bonacci2020contact}
F.~Bonacci, X.~Chateau, E.~M. Furst, J.~Fusier, J.~Goyon, A.~Lema{\^\i}tre, Contact and macroscopic ageing in colloidal suspensions, Nat. Mater. 19~(7) (2020) 775--780.
\newblock \href {https://doi.org/10.1038/s41563-020-0624-9} {\path{doi:10.1038/s41563-020-0624-9}}.

\bibitem{clarke2021gel}
A.~Clarke, Gel breakdown in a formulated product via accumulated strain, Soft Matter 17~(34) (2021) 7893--7902.
\newblock \href {https://doi.org/10.1039/D1SM00816A} {\path{doi:10.1039/D1SM00816A}}.

\end{thebibliography}
\end{document}